%% file: speculation.tex
\PassOptionsToPackage{capitalize,noabbrev,nameinlink}{cleveref}
\PassOptionsToPackage{usenames,dvipsnames}{color}
\PassOptionsToPackage{usenames,table}{xcolor}
\PassOptionsToPackage{final}{microtype}
\PassOptionsToPackage{scaled}{inconsolata}

\documentclass[letterpaper,twocolumn,10pt]{article}
\usepackage{usenix2019_v3}

\newcommand{\paperTitle}{Distributed Speculative Execution for Resilient Cloud
Applications}
\newcommand{\paperKeywords}{Distributed System; Recoverability; Speculative Execution}
\newcommand{\paperAuthors}{Tianyu Li, Badrish Chandramouli, Philip A. Bernstein, Samuel Madden}

\newcommand{\eat}[1]{}
\usepackage{amsthm}

\theoremstyle{definition}
\newtheorem{definition}{Definition}[section]

\hypersetup{
    pdfauthor = {\paperAuthors},
    pdftitle = {\paperTitle},
    pdfkeywords = {\paperKeywords},
    pdfborder={ 0 0 0 }
}

\usepackage{amsmath}
\usepackage{boxedminipage}
\usepackage{xspace}
\usepackage{tabularx}
\usepackage{balance}  
\usepackage{url}

\usepackage[font={small}]{caption}
\usepackage{graphicx}
\usepackage{subfig}
\usepackage[usenames,table]{xcolor}
\usepackage{cleveref}
\usepackage{tabularx}
\usepackage{epsfig}
\usepackage{mfirstuc}

\usepackage[final]{microtype}
\usepackage[T1]{fontenc}
\usepackage{graphicx}
\usepackage{subfig}
\usepackage{soul}
\usepackage{pifont}
\usepackage{listings}
\usepackage{tabularx}
\usepackage{array}
\usepackage{booktabs}
\usepackage[end]{algpseudocode}

\usepackage{enumitem}

\AtBeginEnvironment{thebibliography}{\linespread{1.1}\selectfont}  
\newcommand{\mlbegin}{\shortstack\bgroup}
\newcommand{\mlend}{\egroup}

\newenvironment{squishitemize}{
    \begin{itemize}[leftmargin=*,itemsep=0em]
        \setlength{\parsep}{0pt}
        \setlength{\topsep}{0pt}
        \setlength{\partopsep}{0pt}
        \setlength{\labelwidth}{0.25em}
        \setlength{\labelsep}{0.5em}
    }{\end{itemize}
}

\newcounter{Lcount}
\newcommand{\squishlist}{
    \begin{list}{\arabic{Lcount}. }
   { \usecounter{Lcount}
        \setlength{\itemsep}{0pt}
        \setlength{\parsep}{3pt}
        \setlength{\topsep}{3pt}
        \setlength{\partopsep}{0pt}
        \setlength{\leftmargin}{2em}
        \setlength{\labelwidth}{1.5em}
        \setlength{\labelsep}{0.5em} } }

\newcommand{\squishend}{\end{list}}

\definecolor{todo-color}{rgb}{1,0,0}

\newcommand{\sysname}{libDSE\xspace}

\newcommand{\action}{action\xspace}
\newcommand{\session}{sthread\xspace}
\newcommand{\Action}{Action\xspace}
\newcommand{\Session}{Sthread\xspace}

\definecolor{commentsColor}{rgb}{0.497495, 0.497587, 0.497464}
\definecolor{keywordsColor1}{rgb}{0.000000, 0.000000, 0.635294}
\definecolor{keywordsColor2}{rgb}{0.558215, 0.000000, 0.135316}
\definecolor{bggray}{rgb}{0.97,0.97,0.97}

\lstdefinelanguage{Csharp}{
  keywords={new, void, int, long, bool, byte, return, null, catch, switch, const, if, while, do,
  else, case, break, continue, finally, try, override},
  keywordstyle=\color{keywordsColor1}\bfseries,
  ndkeywords={var, struct, class, interface, true, false, null},
  ndkeywordstyle=\color{keywordsColor2}\bfseries,
  identifierstyle=\color{black},
  sensitive=false,
  comment=[l]{//},
  morecomment=[s]{/*}{*/},
  commentstyle=\color{commentsColor}\textit,
  stringstyle=\color{red}\ttfamily,
  morestring=[b]',
  morestring=[b]"
}

\begin{document}

\title{\paperTitle}

\author{
{\rm Tianyu Li}\\
MIT CSAIL
\and
{\rm Badrish Chandramouli}\\
Microsoft Research
\and
{\rm Philip A. Bernstein}\\
Microsoft Research
\and
{\rm Samuel Madden}\\
MIT CSAIL
} 





\maketitle
\begin{abstract}
Fault-tolerance is critically important in highly-distributed modern cloud applications. Solutions such as Temporal, Azure Durable Functions, and Beldi hide fault-tolerance complexity from developers by persisting execution state and resuming seamlessly from persisted state after failure. This pattern, often called \emph{durable execution}, usually forces frequent and synchronous persistence and results in hefty latency overheads. In this paper, we propose \emph{distributed speculative execution} (DSE), a technique for implementing the durable execution abstraction without incurring this penalty. With DSE, developers write code assuming synchronous persistence, and a DSE \emph{runtime} is responsible for transparently bypassing persistence and reactively repairing application state on failure. We present \sysname, the first DSE application framework that achieves this vision. The key tension in designing \sysname is between imposing restrictions on user programs so the framework can safely and transparently change execution behavior, and avoiding assumptions so \sysname can support more use cases. We address this with a novel programming model centered around message-passing, atomic code blocks, and lightweight threads, and show that it allows developers to build a variety of speculative services, including write-ahead logs, key-value stores, event brokers, and fault-tolerant workflows. Our evaluation shows that \sysname reduces end-to-end latency by up to an order of magnitude compared to current generations of durable execution systems with minimal runtime overhead and manageable complexity. 
\end{abstract}

\input{sections/intro}

\input{sections/bg}

\input{sections/api}
\input{sections/protocol}
\input{sections/discussion}

\input{sections/eval}
\input{sections/rel}

\section{Conclusion}
\label{sec:conclude}
We presented distributed speculative execution (DSE), a novel approach to reduce persistence overhead in durably executed clodu applications. Through our framework, \sysname, we demonstrate that DSE is both practical and beneficial for performance, and that its complexities can be effectively hidden from most developers through the use of abstractions. We built 4 speculative services using \sysname and evaluated 3 end-to-end applications assembled from those building blocks. Our results show that DSE significantly reduces latency. We envision DSE to be a key technique in building future highly distributed cloud applications.

\section*{Acknowledgments}
We are grateful for the support of the MIT DSAIL@CSAIL member companies. We also thank Frans Kaashoek for his helpful comments and suggestions.

\balance

\bibliographystyle{abbrv}
\bibliography{speculation}
\end{document}

%% file: sections/intro.tex
\section{Introduction}
\label{sec:intro}
Modern cloud applications are more distributed than ever. Companies have long adopted microservices architectures, splitting their applications into hundreds of loosely-coupled, independently-deployed distributed components~\cite{deathstarbench, twitter13}.
Recent proposals, such as serverless and granular computing~\cite{jonas2019cloud, lee2019granular}, have pushed for even more distribution at a finer granularity.
Benefits aside, this makes every modern cloud application a distributed one, and fault-tolerance is a key challenge.
A common solution is to build fault-tolerant abstractions and hide failures from the average application developer; such abstractions are responsible for persisting application state where necessary and transparently recover from failures to resume execution. A variety of systems can be thought to fit under this paradigm, including transactional database systems, batch processing systems like Spark~\cite{zaharia12resilient}, exactly-once stream processing systems such as Kafka Streams~\cite{wang21consistency}. Most relevant to the cloud, however, are \emph{Durable Execution}~\cite{durable-execution} systems such as Azure Durable Functions~\cite{df} and Temporal~\cite{temporal}. Compared to previous systems, durable execution engines are designed for the modern cloud and are widely used to orchestrate microservices or serverless workers~\cite{riccomini23}. Despite growing popularity~\cite{zhang2020beldi, jia21boki, zhuang23exoflow, restate, orkes, stealthrocket, littlehorse, flawless, convex, rama, dbos}, current generation durable execution engines perform frequent and synchronous state persistence, which results in high latency overhead (\cref{sec:background}). Worse, this overhead scales with the degree of distribution --- as the application is split into more distributed components, synchronous persistence becomes more frequent and expensive.

In this paper, we propose an alternative: \emph{distributed speculative execution} (DSE), which decouples the abstraction of durable execution from the physical execution pattern. Analogous to similar schemes in hardware~\cite{smith81} and file systems~\cite{nightingale08rethink, nightingale05speculator}, DSE systems bypass synchronous persistence, but ensure equivalence to non-speculative user programs through automatic rollback and recovery after failures. Externally visible results are delayed until persistence to avoid exposing temporary inconsistencies. On the common, failure-free path, DSE dramatically reduces application latency; in exchange, failure recovery becomes slower and more complex. This is a good trade-off when failures are rare, as often the case in the cloud. 

We present \sysname, a \emph{general-purpose} DSE runtime applicable to diverse applications including fault-tolerant workflows, stream processing, and distributed transactions. Compared to earlier work, \sysname is specifically designed for the modern cloud environment where applications are written as workflows over message-passing, fail-restart stateful components. There are two main challenges in designing \sysname. First, to achieve generality, \sysname must perform DSE without relying on application-specific assumptions such as determinism, known read/write sets, or immutability. Second, \sysname must hide the complexities of speculation, but enforce key properties needed for correct DSE in user applications.

For the first challenge, \sysname modifies the 
\emph{Distributed Prefix Recovery} (DPR) protocol~\cite{li21dpr}, a distributed cache recovery scheme, for our setting. DPR instruments messages between services to build and maintain a recovery dependency graph. The graph is used to determine when results are safe to expose externally, and to compute a consistent snapshot for every affected component to rollback to after failure.
Each component allows \sysname to force persistence/rollback when necessary, which requires application changes. 

For the second challenge, \sysname provides a programming model centered around
two intuitive programming primitives, \emph{\action{s}} and \emph{\session{s}}.
\Action{s} are user-defined atomic blocks that are guaranteed by \sysname not to overlap with persist/recovery/rollback invocations.
\Session{s} are stateless and lightweight threads of execution that allow \sysname to guarantee correctness for long-running or asynchronous operations which cannot execute atomically.  
Advanced developers exercise fine-grained control over speculation behavior through \emph{speculation barriers}, analogous to memory barriers~\cite{memorybarriers}.
Barriers allow \sysname services to co-exist with non-speculative components by only exposing non-speculative results.

Using \sysname, developers can build speculative services without worrying about the distributed protocol design and concurrency control challenges of speculation.
These speculative services can be composed into performant yet resilient applications. To showcase this, we build a speculative write-ahead log service, a key-value store, an event broker, and a workflow orchestration service using \sysname. We demonstrate the benefits of DSE by assembling three representative applications from these building blocks: a travel reservation system, an event processing application, and an optimized two-phase commit protocol. Our experiments show that DSE can effectively reduce latency by up to an order of magnitude, with minimal reduction of throughput and scalability. In summary, our contributions are:
\begin{squishitemize}
\item We propose a new distributed speculative execution (DSE) scheme that dramatically and transparently reduces fault-tolerance overhead in durably executed applications.
\item We describe our design and implementation of \sysname, the first general-purpose DSE framework for the modern cloud.
\item We evaluate \sysname on a diverse set of cloud applications, 
 and present a detailed evaluation, showing 20\% to up to an order of magnitude latency savings compared to current, non-speculative implementations. 
\end{squishitemize}

%% file: sections/bg.tex
\section{Motivation}
\label{sec:background}

\subsection{Background: Durable Execution}
Durable execution creates an illusion of uninterrupted, failure-free execution by persisting application state at every step and automatically retrying execution after failure. This strong abstraction hides many distributed system complexities from the average developer, but often at the cost of performance.

\vspace{.1in}
\noindent\textbf{Running Example.}
Consider a toy example of an application that consists of a frontend layer, a counter service, and a logging service. Users issue requests to the frontend to increment the counter. Each request and its result is then sent to the logging service for bookkeeping purposes. Failures can result in a number of anomalies in even this simple scenario. Each service can crash and lose data. The frontend server can crash after a successful increment but fail to log the incremented value. Naive attempts to solve this may retry and increment or log too many times. While these anomalies appear innocuous in the toy example, they are highly undesirable in real-world scenarios. If an e-commerce company uses a similar architecture to connect their payment and fulfillment services, such anomalies can lead to double charging or unfulfilled orders. A principled solution is necessary to prevent such anomalies.

\begin{figure}[t!]
  \centering
  \includegraphics[width=\columnwidth]{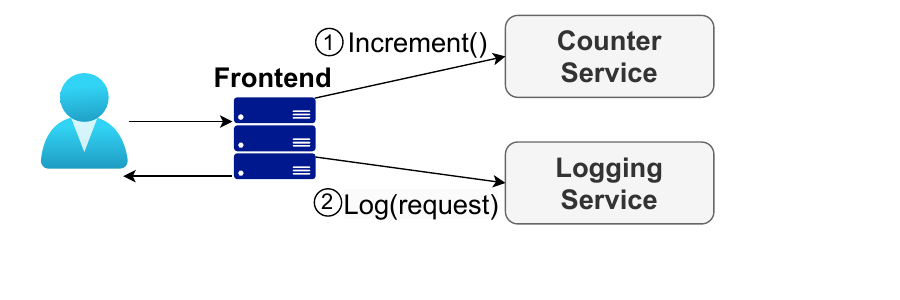}
  \caption{A Simple Running Example}
  \label{fig:arch}
\end{figure}

\vspace{.1in}
\noindent\textbf{Current Solutions.}
The goal of durable execution is to ensure that users cannot distinguish between execution of the application with failures, and ones without (except perhaps through performance degradation).
Today's cloud architect might achieve durable execution in this example by first making each service durable (e.g., by persisting to a database before returning), and then using a resilient workflow engine (e.g., Temporal) to orchestrate the frontend calls. Importantly, each stateful service must be programmed for \emph{idempotency} -- retried requests must not have additional visible effects beyond their first execution. Often this is achieved by attaching a unique ID to each request and de-duplicating on the service side. To protect against frontend failure, the workflow engine automatically persists operation intents~\cite{setty16olive} and intermediate results, and retries on failure. In our running example, the workflow engine would execute an increment call by first persisting the intent to increment and generate a request ID, make the remote call, and record the result.

Under the hood, the workflow system maintains (sometimes implicitly) a directed acyclic graph (DAG) of stateful tasks, and ensures exactly-once semantics of execution (assuming idempotent tasks). Most such systems today must synchronously persist each task's output before starting the next task to ensure correctness. Otherwise, because tasks might not be deterministic, replay may yield different outputs and spawn different downstream tasks; tasks spawned based on previous outputs can then conflict with later ones and cause anomalies. 
Consequently, almost all current generation durable execution systems have cumulative latency overheads that scale with the depth of the task graph. State-of-the-art implementations such as Boki~\cite{jia21boki} and Halfmoon~\cite{qi23halfmoon} focus on optimizations that increase throughput and reduce overhead per-persistence, but do not fundamentally bypass synchronous persistence. ExoFlow~\cite{zhuang23exoflow} relies on annotations to mark tasks that can be deterministically replayed or with effects that can be rolled back to avoid synchronous persistence. However, this solution relies on users to carefully annotate each task and supply the logic to rollback outputs, which increases the burden on the  developer. This approach also does not help with non-deterministic applications.

\vspace{.1in}
\noindent\textbf{Our Solution.}
In contrast, DSE forges ahead before operation intents, effects, or intermediate results are durable. This inevitably leads to anomalies upon failures. DSE applications must detect such inconsistencies, repair them, 
and prevent external users from observing them during the process. In other words, the requirements of correct DSE are:
\begin{squishitemize}
\item \emph{Recoverability.} Inconsistencies are eventually repaired.
\item \emph{Failure Transparency.} External entities (e.g., clients) never observe failure-induced inconsistencies.
\end{squishitemize}
Fundamentally, this means that DSE systems are rollback-based recovery systems~\cite{elnozahy02survey}. Any solution must design protocols that correctly track recovery dependencies between participants, determine when state becomes recoverable, and rollback unrecoverable state on failure. On failure, the system determines the extent of failure based on collected dependencies and orchestrates rollback across participants to restore the application to a consistent state. Such schemes support non-determinism, as non-deterministic operations are either restored from persisted state and never replayed, or all their effects are rolled back. The key trade-off is that on failure, DSE throws away work and requires rollback coordination, which adds overhead. In exchange, DSE optimizes for latency by removing synchronous persistence from the critical path. Overall persistence overhead of a DAG is no longer the sum of all steps, but instead is the maximum of all steps, as asynchronous persistence occurs in parallel.

\subsection{Example Applications}
Next, we sketch how DSE applies to three classes of applications to improve end-to-end latency. In all of these cases, correctness requirements force existing implementations to frequently and synchronously persist state, causing increased latency that DSE can help address. We later describe how we build an application from each class with \sysname and demonstrate the benefits in \cref{sec:eval}.

\begin{figure}[t!]
  \centering
  \includegraphics[width=0.9\columnwidth]{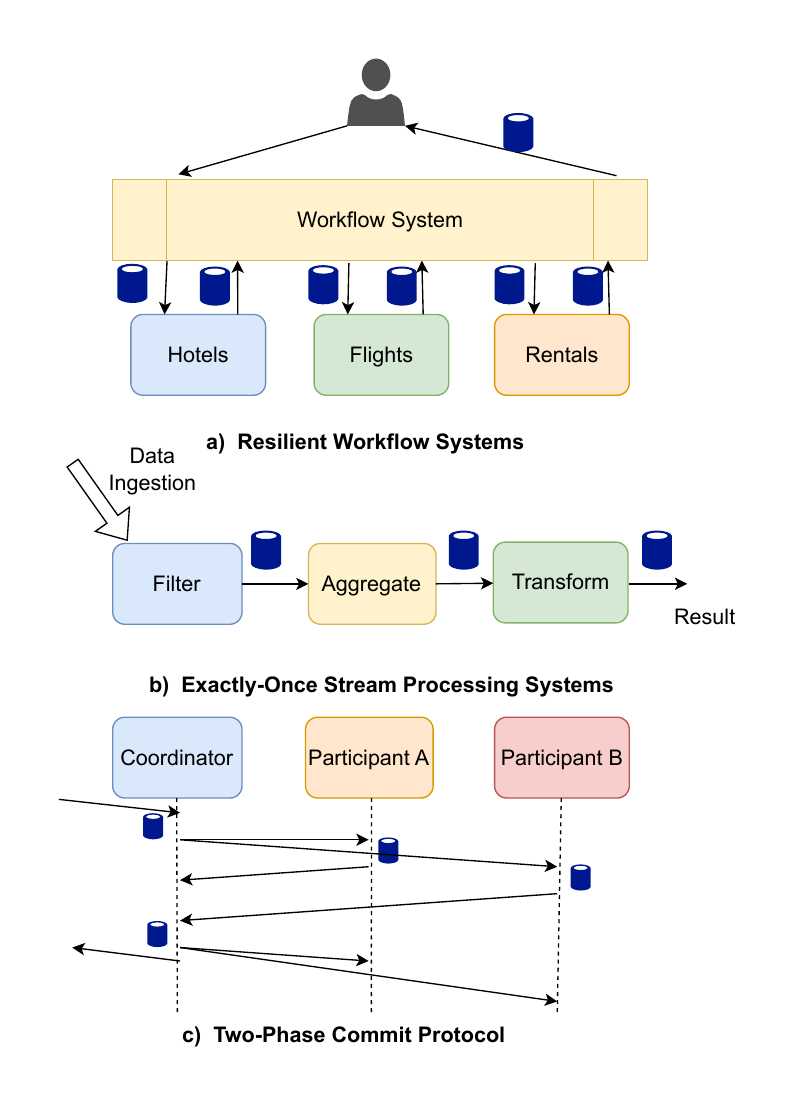}
  \caption{Example Applications that benefit from DSE}
  \label{fig:examples}
\end{figure}

\noindent\textbf{Resilient Workflows.} As mentioned, resilient workflow systems typically must persist their state synchronously between steps of the workflow. In the example of \cref{fig:examples}a, a travel-reservation application uses a system like Temporal to orchestrate reservations between three independent services. The workflow system must persist the intent to make a reservation, wait for the reservation to become persistent, and persistently record the outcome before proceeding, leading to significant synchronous persistence.
\vspace{.1in}

\noindent\textbf{Event-driven Processing.}
Many modern cloud systems consist of decoupled services that react to events, and rely on event broker (pub-sub) systems~\cite{kreps2011kafka, eventhubs} for inter-service communications. Event brokers must be stateful and fault-tolerant to ensure availability, support asynchronous/unreliable consumers, etc. Most brokers therefore disallow event consumption before persistence. In \cref{fig:examples}b, a stream processing application (e.g., Kafka Streams~\cite{wang21consistency}) relies on an event broker to pass data between operators, and therefore must pay one synchronous persistence per operator. 
\vspace{.1in}

\noindent\textbf{Distributed System Primitives.} 
Critical distributed protocols often involve synchronous persistence for fault-tolerance. For example, as shown in \cref{fig:examples}c, the vanilla two-phase commit~\cite{mohan86transaction} protocol requires the transaction coordinator to log start of commit before starting the protocol, workers to log their responses before responding to PREPARE, and the transaction coordinator to log its decisions of commit/abort before completing the protocol and notifying participants. DSE provides a new way to transparently optimize such protocols, bypassing synchronous persistence to the log without the need to redesign the protocol. 

%% file: sections/api.tex
\section{\sysname API}
\label{sec:api}

\begin{table*}[ht]
\centering 
\captionsetup{justification=centering} 
\rowcolors{2}{gray!25}{white} 
\begin{tabular}{
>{\ttfamily}p{0.31\textwidth}
>{\sffamily}p{0.68\textwidth}
}
\hline
\textbf{\texttt{StateObject} abstract API} & \textbf{Semantics} \\ 
\hline
Persist(v, metadata, callback) & Persist the current state and \texttt{metadata} as version \texttt{v} and then invoke \texttt{callback}. \\
Restore(v) $\rightarrow$ metadata & Recover (or rollback) to the given version and return associated metadata. \\
Prune(v) & Signal that a version (and all that precede it) can be pruned. \\
ListVersions() $\rightarrow$ metadata[] & List all unpruned versions and associated metadata. \\
\hline
\end{tabular}
\caption{\texttt{StateObject} Abstraction} 
\label{table:stateobject} 
\end{table*}

\begin{table*}[ht]
\centering 
\captionsetup{justification=centering} 
\rowcolors{2}{gray!25}{white} 
\begin{tabular}{
>{\ttfamily}p{0.3\textwidth}
>{\sffamily}p{0.65\textwidth}
}
\hline
\textbf{\texttt{StateObject} API} & \textbf{Semantics} \\ 
\hline
StartAction(header) $\rightarrow$ bool & Start an atomic \action, (optionally) receive 
the given \sysname message header. If returned false, the message originates from rolled back state and should be discarded.\\
EndAction() $\rightarrow$ header & End an atomic \action and (optionally) produce a header that encodes dependencies for any outgoing messages resulting from this \action. \\
Connect(config) & Connect this \texttt{StateObject} to a \sysname Coordinator using the given configuration. Must be invoked exactly once before performing any other operation. \\
Refresh() & Refresh should be periodically called to perform \sysname background tasks and guarantee progress for the cluster. \\
Detach() $\rightarrow$ \session & Temporarily suspend an \action and produce an \session that can be used for speculative operations independent of this \texttt{StateObject}.\\
Merge(\session) $\rightarrow$ bool & Resume an \action by merging the detached \session backed into this \texttt{StateObject} and update its dependencies. If false, the state leading to the creation of the \session was rolled back and the \session should be discarded.\\
\hline
\textbf{\texttt{\session} API} & \textbf{Semantics} \\ 
\hline
Receive(header) $\rightarrow$ bool & Attempt to receive the given \sysname message header. If returned false, the message should be discarded. Throws exception if the \session is rolled back. \\
Send() $\rightarrow$ header & Produce a header that encodes current dependencies of this \session. \\
Barrier() & Blocks until all state observed by this \session becomes non-speculative. Throws exception if the \session is rolled back. \\
\hline
\end{tabular}
\caption{Summary of \sysname API} 
\label{table:api} 
\end{table*}

DSE application developers build speculative services by encapsulating application state into \texttt{StateObject}s. \texttt{StateObject}s are stateful, message-passing entities loosely based on I/O automata~\cite{lynch87hierarchical} that follow a \emph{fail-restart} model~\cite{li23darq}. Each \texttt{StateObject} can persist its state and recover to it after failure, thereby appearing to restart instead of failing. 
Application developers define \texttt{StateObject}s by implementing a set of methods to persist or restore application state, using \sysname's primitives to safely access and update that state. Application developers do not reason about speculation or failure recovery, as it is the responsibility of the \sysname runtime to 1) persist, recover, or rollback \texttt{StateObject}s when necessary by invoking developer-supplied methods, 2) instrument application messages to establish dependencies, drop rolled back messages, or delay speculative messages from external observers, and 3) implement the necessary protection when developers access \texttt{StateObject}s through \sysname provided primitives. The rest of this section focuses on these primitives (\cref{table:api}). We discuss the details of how \sysname orchestrates speculative execution in \cref{sec:protocol}. 
\vspace{.1in}

\subsection{StateObjects and Actions}
Users create \texttt{StateObject}s by implementing the abstract API shown in \cref{table:api}. We show an implementation of the counter service from our running example using cloud storage files in \cref{alg:example-so}. The framework ensures that \texttt{Persist} and \texttt{Restore} are not concurrently invoked. For performance, the \texttt{Persist} API is asynchronous; users may return from the call once the persistence operation is issued (e.g., flush the log at a specific offset), but before it is completed. While conceptually simple, \texttt{Restore} encapsulates two different use cases, one where the \texttt{StateObject} has failed and must reload state from persistent storage, and another where the \texttt{StateObject} responds to a rollback. 
For the latter case, the example loads persistent state. This is correct, but unnecessarily slow and inefficient.
Developers are therefore encouraged to distinguish between these two cases (e.g., by checking if \texttt{value == default} in the example) and apply application-specific optimizations if possible (e.g., by leveraging built-in multi-versioning~\cite{reed78mvcc, berenson95critique, prasaad19cpr}).
If the \texttt{StateObject} crashes, the runtime uses \texttt{ListVersions} to determine (unpruned) successful \texttt{Persist} calls. This API easily allows for alternative implementations of persistence and recovery. For example, one can implement the counter service via either logging and replay or a replicated state machine.

\begin{figure}[t!]
    \input{sections/code/example-stateobject}
    \caption{
        \textbf{Example \texttt{StateObject} Implementation}
    }
    \label{alg:example-so}
\end{figure}

\begin{figure}[t!]
    \input{sections/code/example-service}
    \caption{
        \textbf{Example CounterService Implementation}
    }
    \label{alg:example-service}
\end{figure}

\sysname organizes operations on \texttt{StateObject}s as a series of atomic execution units called \action{s}. Each action's effects are either entirely persisted or entirely lost due to recovery or rollback. \sysname achieves this by ensuring \action{s} never interleave with \texttt{Persist} or \texttt{Restore} operations. However, \action{s} may execute concurrently with each other to allow for parallelism. Messages are consumed as a part of \action{s} or produced as a result of \action{s}. \sysname provides users with flexibility on how best to deliver messages between \texttt{StateObject}s, relying on users to pass opaque headers instead of implementing messaging in \sysname. We present an example of this in \cref{alg:example-service}. Here, we implement our counter service using gRPC~\cite{grpc}, and declare a metadata field on each message for \sysname headers. To process an increment request, the developer begins an \action by consuming the input header.  If \texttt{StartAction} returns false, the message's sender has been rolled back, and the message must be discarded. Otherwise, \sysname protects execution after line 21 from \sysname-triggered persists and restores. Users still need to guard against multiple concurrent requests, and hence use the thread-safe version of increment in line 23. Finally, developers end the action, obtaining a header to pass back to the caller. 

\subsection{Handling Asynchrony}
Long running \action{s} can block persistence until finished and halt persistence progress across the cluster. For example, the frontend server in our running example issues RPCs that may complete asynchronously and automatically retry in the background. \sysname introduces \emph{\session{s}} to support such use cases. An \session is essentially a lightweight thread of execution within a \texttt{StateObject} that encodes the speculative state of its parent \texttt{StateObject} at time of \session creation. We show an example in \cref{alg:example-advanced}. Here, the frontend workflow orchestrator calls \texttt{Detach} after logging the intent to call counter service (line 6), but before calling (line 10).
After line 7, application code no longer runs in an atomic block, and may go to sleep, retry, or otherwise perform long-running operations without blocking the \texttt{StateObject}.
Consequently, in the event of a rollback, the \session may still temporarily continue to execute. It is therefore essential that \session{s} are treated as \emph{independent, standalone} participants of the system and interact with other participants, including its parent \texttt{StateObject}, exclusively through \sysname-instrumented message passing. For example, in line 9, we generate \sysname headers from the \session, and consume headers on line 12. Finally, to update the parent \texttt{StateObject}, \session{s} must be merged to start a new \action (line 13). This logically sends a message from \session to \texttt{StateObject}, which can be rejected; in this case, the \session simply terminates, as it represents derived state of its parent, which will recover independently.

\begin{figure}[t!]
    \input{sections/code/example-advanced}
    \caption{
        \textbf{Example Workflow Implementation with \sysname}
    }
    \label{alg:example-advanced}
\end{figure}

Finally, \session{s} support barriers, which provide fine-grained control over speculative behavior. Similar to a memory fence, a barrier blocks until everything the \session received becomes non-speculative, thus preventing speculation across the barrier. Only \session{s} are allowed to invoke barriers, as they are, by definition, blocking. Barriers are useful when interacting with external entities. For example, in line 19 and 20 of \cref{alg:example-advanced}, by detaching and calling barrier, line 21 will be delayed until the response becomes non-speculative, therefore only sending non-speculative results to users. In many cases, barriers can be automatically inserted (e.g., at the end of an RPC handler processing an external user request), but advanced users may also use barriers to prevent speculative dependencies between parts of their applications. For example, a barrier before line 13 prevents the logging service from receiving speculative requests.

%% file: sections/code/example-stateobject.tex
\begin{lstlisting}[numbers=left,
      numberstyle=\small,
      numbersep=6pt,
      xleftmargin=2em,
      xrightmargin=0.5em,
      framexleftmargin=1.5em,
      frame=single]
class CounterStateObject : StateObject {
  int value;
  
  void Persist(long ver, byte[] m, Action c) {
    var content = Encode(ver, m);
    // async write to a cloud storage
    Task.Run(() => {
      WriteFile(GetName(v), content);
      callback();
    });
  }

  byte[] Restore(long ver) {
    content = ReadFile(GetName(ver));
    var restored, metadata = Decode(content);
    value = restored;
    return metadata;
  }
  ...
}
\end{lstlisting} 

%% file: sections/code/example-service.tex
\begin{lstlisting}[numbers=left,
      numberstyle=\small,
      numbersep=6pt,
      xleftmargin=2em,
      xrightmargin=0.5em,
      framexleftmargin=1.5em,
      frame=single]
message IncrementRequest {
  bytes header = 1;
  int incrementBy = 2;
}
message IncrementResponse {
  bytes header = 1;
  int result = 2;
}
service CounterService {
  ...
}

// gRPC generated interface
class CounterImpl : CounterServiceBase {
  CounterStateObject so;

  override IncrementResponse Increment(
      IncrementRequest r) {
    if (!so.StartAction(r.header)) throw;
    response = new IncrementResponse();
    response.result =
        AtomicInc(so.value, r.incrementBy);
    response.header = so.EndAction();
    return response;
  }
  ...
}
\end{lstlisting} 

%% file: sections/code/example-advanced.tex
\begin{lstlisting}[numbers=left,
      numberstyle=\small,
      numbersep=6pt,
      xleftmargin=2em,
      xrightmargin=0.5em,
      framexleftmargin=1.5em,
      frame=single]
...
WorkflowEngineStateObject so;
...
async Task ExecWorkflow(WorkflowRequest r) {
  if (!so.StartAction(r.header)) throw;
  so.MarkWorkflowStart(r);
  var  t = so.Detach();
  // Request executes outside atomic block
  cReq = ComposeRequest(r, t.Send());
  var cRes =
      await CallCounterServiceAsync(cReq);
  if (!t.Receive(cRes.header)) throw;
  if (!so.Merge(t)) throw;
  // Back in an atomic block
  so.UpdateWorkflow(cRes);
  ...
  // Finally, return after barrier to
  // hide speculation from caller
  t = so.Detach();
  await t.Barrier();
  return response;
}
\end{lstlisting} 

%% file: sections/protocol.tex
\section{Speculative Protocol}
\label{sec:protocol}

In this section, we describe how \sysname achieves DSE based on our application model of message-passing atomic \action{s}.

\subsection{Inspiration: DPR}
\begin{figure}[t!]
  \centering
  \includegraphics[width=\columnwidth]{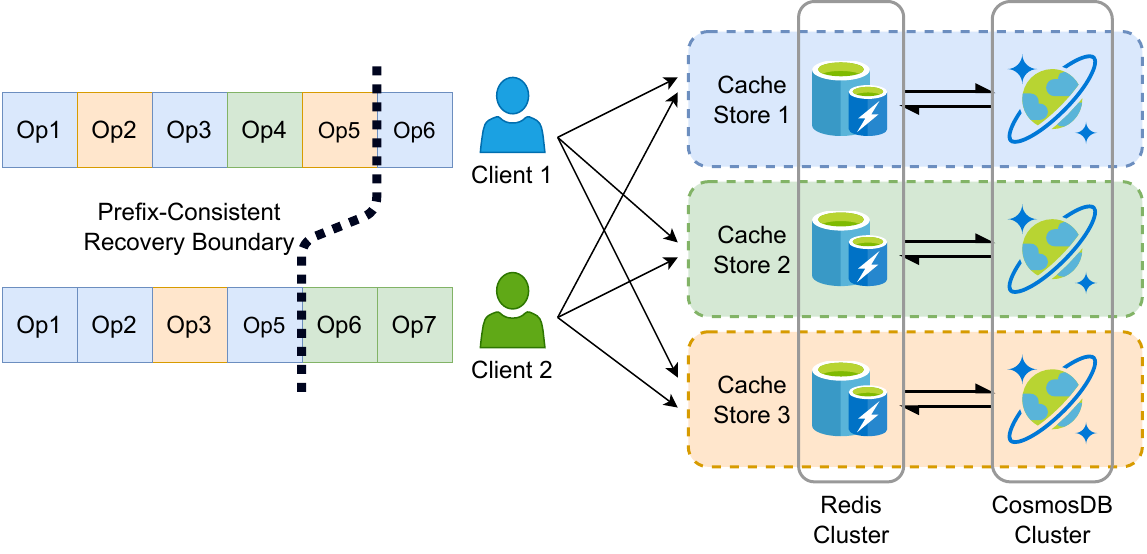}
  \caption{Distributed Prefix Recovery}
  \label{fig:dpr}
\end{figure}

Distributed Prefix Recovery (DPR)~\cite{li21dpr} is a recently proposed technique for causal consistency across sharded ``cache-stores'',
storage units spanning volatile memory and durable storage as shown
in \cref{fig:dpr}. For example, a shard within a partitioned database with an accompanying Redis caching layer can be considered a (logical) cache-store. Often in such an architecture, the caches are \emph{write-through}: clients read from the cache, but write directly to the backend storage and ensure consistency with invalidation. In contrast, DPR is \emph{write-back}: writes directly update the cache for increased throughput and immediate visibility, and cached entries are asynchronously flushed to storage. This leads to lost writes if caches fail, which may cause application anomalies as readers may have acted on lost updates. DPR presents a lightweight protocol for addressing this problem. First, operation completion and persistence are decoupled using two acknowledgements. Second, DPR clients interact with cache-stores explicitly through \emph{sessions}. Each session is a (linearizable) sequence of operations, as shown in the left of \cref{fig:dpr}, and DPR guarantees, through orchestrated rollbacks of state, that any surviving state corresponds to session \emph{prefixes}. This ensures that in the event of a cache failure, no surviving operation depends on a lost operation.

\subsubsection{Our Insights}
The original formulation of DPR around cache-stores does not easily apply to our setting. In particular, clients are first-class citizens in DPR, but do not participate in rollbacks; they are responsible for discovering rollbacks and resuming (potentially replaying) operations. This complexity is compounded by possible client failures, as the recovered (amnesiac) client may lack information needed to properly handle a rollback. Unlike DPR's client-server formulation, \sysname's speculative protocol is defined between message-passing stateful peers and makes no distinction between requests and responses, leading to a more succinct protocol than DPR. One benefit of this formulation is that application control flow is part of persisted state (e.g., as part of the workflow orchestrator) and will therefore participate in any rollback, allowing applications to resume from exactly where they are supposed to without needing to recover clients separately. 

We improve upon DPR by introducing a new coordinator design. The original DPR relies on a stateful coordinator (i.e., the point of truth is the persistent state of the coordinator) built on an external database system for correct operations. Consequently, DPR adds one coordinator persistence to the failure-free code path before writes can be declared recoverable. We present a deterministic and stateless coordinator design (i.e., the point of truth is the collective persistent state of participants), which removes this overhead.

\subsection{Protocol Details}
\label{sec:protocol-details}
\begin{figure}[t!]
  \centering
  \includegraphics[width=0.6\columnwidth]{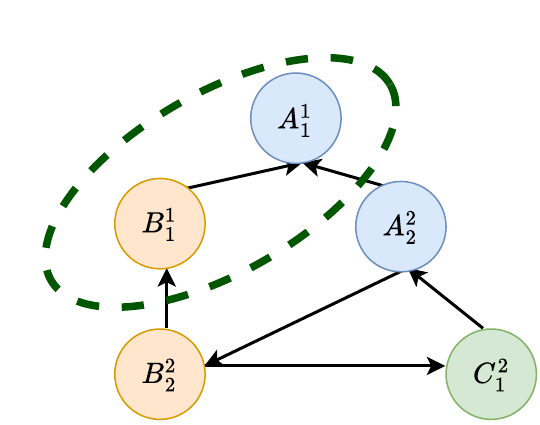}
  \caption{Example Dependency Graph}
  \label{fig:dep}
\end{figure}

Like DPR, the \sysname protocol centers around explicit recovery dependency tracking with a dependency graph, as shown in \cref{fig:dep}. Each vertex of the graph is a recoverable point that can be loaded via \texttt{Restore}, and is uniquely identified by a combination of a \texttt{StateObject} id, a global failure counter (more details on this later), and a local persistence counter. For example, $A^1_2$ is the recoverable point from \texttt{StateObject} A with local persistence counter 2 and global failure counter 1. Edges of the dependency graph represent recovery dependencies -- an edge from $u$ to $v$ iff $u$ recovering without $v$ results in an inconsistency. By definition of consistency~\cite{lamport78time}, such a dependency is established either implicitly by precedence (i.e., $u$ is later than $v$ in the same \texttt{StateObject}), or if $u$ received a message originating from state captured by $v$. Because each recoverable point captures multiple state transitions and messages, the dependency graph may have cycles.

\noindent\textbf{Instrumentation Protocol.}
\sysname tags each message with its originating vertex on the dependency graph (the content of aforementioned \sysname headers). If a \texttt{StateObject} receives a message from $v$ when its current (not yet persisted) recoverable point is $u$, it adds an edge from $u$ to $v$ in the dependency graph. Each \texttt{StateObject} accumulates dependency graph updates and sends them periodically to the \sysname coordinator after persistence. For \session{s}, because they are derived state, they do not participate on the dependency graph, but track all of their dependencies in a set and append them all to each message. The dependency set of an \session can be cleared every time after a barrier to prevent unbounded growth.

\noindent\textbf{Boundary Protocol.}
A \emph{Recoverable Boundary} is a cut of the graph where no future failure will cause vertices within the boundary to roll back. Messages originating from within the boundary are therefore not speculative and safe to expose externally. Recoverable boundaries manifest on the graph as closures, i.e., sets of vertices that 1) are all recoverable and 2) have no edge leading to non-recoverable vertexes, as seen in \cref{fig:dep}. The coordinator finds boundaries by periodically searching the graph. To ensure the existence of closures, we additionally require the following:
\begin{definition}{(Commit Ordering Rule)} $A^x_y$ is allowed to receive a message from
	$B^m_n$ only if $y \geq n$.
\end{definition}
\noindent This ensures that for any vertex $A^x_y$, the set of vertices with persistence number $\leq y$ is a closure that includes $A^x_y$. This modification prevents degenerate cases where a failure may roll back arbitrarily many vertices (i.e., the domino effect~\cite{elnozahy02survey}). The \sysname runtime blocks an \action from starting if the message it receives violates the commit ordering rule until local persistence has caught up. Note that this will not lead to deadlock situations, as local persistence can always advance in our system without depending on other participants.

\noindent\textbf{Recovery Protocol.}
Failures lead to vertex loss on the dependency graph, and the system state is inconsistent iff any surviving vertex has an edge into a lost vertex. To recover, the system must roll back vertices until no such violations exist. The primary challenge here is that the system must achieve \emph{consensus} on rollbacks; otherwise, overlapping failures or out-of-sync view of the dependency graph may result in different parts of the system recovering to different closures and creating inconsistency in the process. For simplicity, our protocol relies on the coordinator (which embodies consensus as a ``leader'' of the cluster) to make a unilateral decision. As part of this consensus, the coordinator also assigns a monotonically increasing and unique failure sequence number to the rollback (i.e., the global failure counter from before). Each participant of the rollback then recover to the prescribed state at their own pace. This creates a temporary split of the cluster into pre-recovery and post-recovery halves. \sysname
identifies communications between them using the different global failure counter values messages carry, and disallows such communication for safety. Additionally, each \texttt{StateObject} recovers strictly in order of assigned failure sequence number. To summarize:
\begin{definition}{(Recovery Sequencing Rule)} $A^x_y$ is allowed to go to $A^{x+1}_z$ if and
	only if $A$ successfully completes the rollback request for failure sequence number $x$. 
\end{definition}
\begin{definition}{(Recovery Partition Rule)} $A^x_y$ is allowed to receive a message from
	$B^m_n$ only if $x = m$. If $m < x$, the message is discarded; otherwise, its 
        receipt must be delayed.
\end{definition}

\noindent\textbf{Correctness Sketch.} Because our protocol is mostly a restatement of the DPR protocol in a message-passing model, its correctness largely follows from DPR (we discuss our new coordinator design in the next subsection). What is left then is to demonstrate that our model and the DPR cache-store model are equivalent. On a high-level, each \sysname \texttt{StateObject} can be thought of as the combination of a cache-store and an execution thread that allows the \texttt{StateObject} to update the cache-store in the absence of external client requests. The execution thread then is logically just a co-located DPR client that always reads the latest cache-store state before issuing a write operation to the cache-store. Message-passing can be modeled as issuing a request to a remote cache-store from this co-located DPR session. Then, one can demonstrate that after simplifications (because the DPR session only reads from the local cache-store), the \sysname protocol is equivalent to the DPR protocol.

\subsection{Coordinator Design}
\begin{figure}[t!]
  \centering
  \includegraphics[width=\columnwidth]{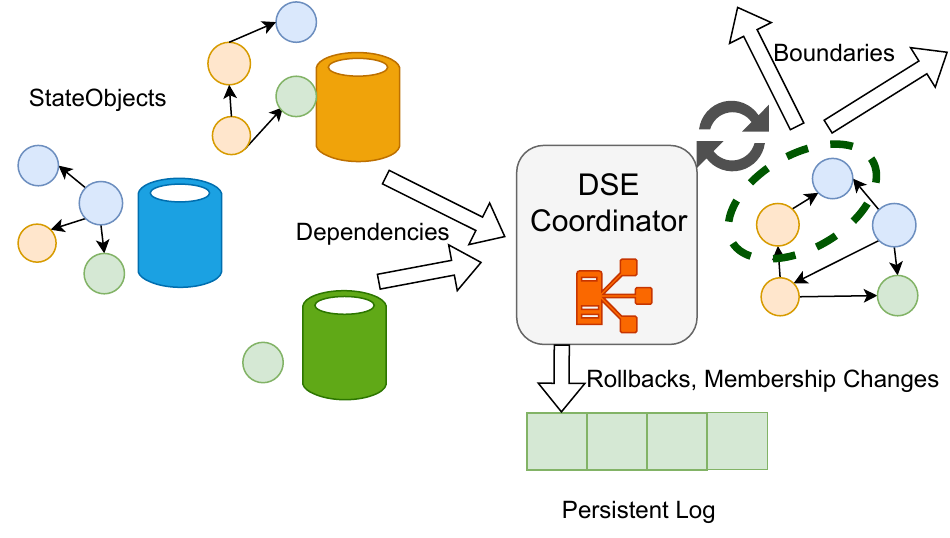}
  \caption{\sysname Coordinator Design 
  }
  \label{fig:coordinator}
\end{figure}

We now describe the design of our coordinator. As shown in \cref{fig:coordinator}, the \sysname coordinator, unlike DPR's, is backed by a persistent log (either through distributed consensus like Raft~\cite{ongaro14raft} or reliable cloud storage) that encodes changes to the cluster state, such as membership changes and recoveries.

\textbf{Finding Boundaries.} Unlike DPR, \sysname automatically persists fragments of the dependency graph as part of \texttt{StateObject}s using the metadata parameter on \texttt{Persist}. This forms a distributed point of truth for the dependency graph, and the coordinator maintains a (outdated) view of the actual graph. Newly persistent versions and in-flight operations may be missing from the coordinator's view. That said, the persistent part of the graph is \emph{immutable} -- future operations may add vertices, but cannot change dependencies in the past. This makes it safe for the coordinator to declare recoverable boundaries on the coordinator's view, because any recoverable boundaries the coordinator finds on its present view must also be recoverable on a later view. Consequently, the coordinator does \emph{not} persist its decisions on the failure-free path -- restarted coordinators instead recompute the boundary on a more up-to-date view to find the same (or larger) recoverable boundary.

\textbf{Orchestrating Rollback.} A rollback is triggered when one of the \texttt{StateObject}s fail, and the coordinator computes the extent of data loss by removing any vertices on the dependency graph not reported as persistent by the restarting \texttt{StateObject}.
Then, it iteratively removes vertices from the graph until no surviving vertex has a dangling edge. As discussed in \cref{sec:protocol-details}, on failure, the coordinator must achieve \emph{consensus} between \texttt{StateObject}s on what the final state of the graph is in response to the restart. This is achieved by synchronously persisting the failure (by assigning it a failure sequence number) and the rollback actions proposed, before releasing the decision to the rest of the cluster on the log. 

\textbf{Coordinator Recovery.} Even though the coordinator is stateless, when a coordinator fails, the system must still perform some necessary steps to ensure correctness. On recovery, the new coordinator needs to reconstruct the cluster state, including previous rollback decisions and the list of active participants by replaying the persistent log. It then re-sends recent rollback decisions and asks for every participant to send locally stored dependency graph segments, which guarantees a view of the dependency graph that is more up-to-date than before the failure. The coordinator cannot answer
queries about the current recovery boundary or before every participant has responded to ensure a complete view of the graph.

%% file: sections/discussion.tex
\section{Implementation and Discussion}
\label{sec:discussion}

\subsection{Putting it Together: \sysname Runtime}
The \sysname runtime implements the aforementioned programming model, and connects each \texttt{StateObject} to the coordinator we just presented. In total, the core \sysname and coordinator implementation consists of around 4000 lines of C\# code. 

\noindent\textbf{Framework Integrations.} \sysname has built-in integration with gRPC~\cite{grpc} and ASP.NET~\cite{aspnet}. To create a speculative service, users first write the service API using vanilla gRPC, encapsulate the state with a \texttt{StateObject}, and implement the generated gRPC stub using the \sysname API. \sysname provides an optional gRPC interceptor~\cite{interceptor} that automatically injects \action boundaries and barriers where appropriate before and after RPC handlers execute, and uses the HTTP headers to transparently pass \sysname headers, hiding speculation from the average developer altogether. Advanced users can disable the interceptor and manually call \sysname APIs where necessary. Finally, all background logic of \sysname is packaged as a ASP.NET managed service, and users can start an ASP.NET-based speculative service simply by declaring these dependencies and leaving the rest to the ASP.NET framework.



\noindent\textbf{Synchronization.} \sysname uses biased locking~\cite{russell06, dice19bravo, li22epvs, li24epvs} to implement \action{s} and ensure scalability. Essentially, each \action executes under shared lock while \texttt{Persist} and \texttt{Restore} execute under exclusive lock.

\noindent\textbf{\texttt{StateObject} Life Cycle.}  When starting an application,  \texttt{StateObject}s first \texttt{Connect} to a the coordinator to report its presence with a predetermined unique ID. If a \texttt{StateObject} that has already reported starting does so again, \sysname treats this as indicative of a failure. This triggers cluster level recovery which sends each affected runtime node a request to recover to a specific point. This assumes that external systems (e.g., Kubernetes) detect down services, replace them, and attempt to reconnect, but this is not a fundamental restriction. When \texttt{Connect} returns, \sysname has registered the calling \texttt{StateObject} as the legitimate incarnation which may start sending and receiving messages. \sysname guarantees that any messages received from or sent to the previous incarnation will eventually be rolled back, but it is up to service implementers to ensure that two incarnations do not simultaneously update persistent state. It is not necessary for a service to explicitly disconnect from a \sysname coordinator, as long as it persists all of its outstanding state before becoming inactive.

\subsection{Building Speculative Services}
We implement four cloud building blocks using \sysname:

\noindent\textbf{Speculative Log.} We implement a speculative write-ahead log based on the open-source FasterLog~\cite{fasterlog} project, and it serves as the basis for our implementation of other speculative services. The most significant change required from a standard log implementation is attaching custom metadata with \texttt{Persist} calls. We achieve this by designing special commit records that hold metadata, and appending them to the end of a log at the beginning of \texttt{Persist}. Recovering or rolling back simply drops all log entry after the latest surviving commit record. Our speculative wrapper maintains an in-memory mapping of (unpruned) commit records and their offsets within the log to speed up this operation, and rollbacks are directly done using this mapping. The mapping is persisted as metadata to avoid scanning the log on recovery. Overall, the speculative log is about 200 lines of wrapper code around FasterLog and took around 2 person-hours to complete.

\noindent\textbf{Key-Value Store.} We implement a speculative key-value store based on the open-source FASTER Key Value store~\cite{faster}, which powers production systems such as Orleans~\cite{orleans}, Durable Functions~\cite{df}, and Garnet~\cite{garnet}. We use \sysname's gRPC integration to add request handling to FASTER, originally an embedded key-value store. Our speculative implementation, including stored-procedures for our workload (discussed later), is about 400 lines of C\# code. 

\noindent\textbf{Speculative Workflows.} We then build a speculative workflow orchestration engine, similar to Durable Functions and Temporal. Our implementation follows the CReSt model used in DARQ~\cite{li23darq}, and achieves speculative durable execution by performing atomic transitions of state on a speculative persistent log. Note here that our implementation only includes the core backend logic of workflow engines, and omit the front-end code-as-workflow infrastructure and supporting features such as durable timers. Overall, the speculative workflow engine contains about 200 lines of C\# code.

\noindent\textbf{Event Broker.} Lastly, we implement a speculative event broker system similar to Kafka~\cite{kreps2011kafka} and EventHubs~\cite{eventhubs}, with a collection of producers and consumers interacting through disjoint topics. Each topic is backed by a collection of partitions, implemented with speculative logs. We build the event broker service again using gRPC, but with manual \sysname integration to allow for streaming gRPC calls. Our broker uses the DARQ API~\cite{li23darq} for exactly-once processing. Building the system took about 10 person-hours and around 800 lines of C\# code.

\subsection{Limitations}
Both the original DPR and our modified protocol rolls back more state than necessary in the case of a failure -- in-flight messages and concurrent operations may introduce new dependencies not yet visible to the coordinator, and without additional protocols, the coordinator must draw the rollback boundary based on the immutable, persistent portion of the graph, which results in all participants rolling back their latest in-memory content. In practice, we mitigate this issue by skipping rollbacks of a local version if a \texttt{StateObject} does not depend on any uncommitted state. This ensures, for example, that participants not exposed to speculative state are never impacted by rollbacks. For further optimization, participants need to vote on the rollback boundary on failure instead; we leave this for future work. The primary issue with \sysname and the DSE model is that existing applications cannot transparently benefit from these techniques. Developers must switch to speculative building blocks or implement their own to get the most out of DSE. That said, the migration can be incremental, as \sysname is backward-compatible with existing services via barriers. The DSE framework and protocol is not designed for replicated services; while simple state machine replications approaches~\cite{schneider90smr} can be thought of as an alternative persistence backend in \texttt{StateObject}s, quorum-based systems or systems with weaker guarantees (e.g., Dynamo~\cite{decandia07dynamo}) do not have a clear ``restart'' semantic and therefore are not easily modeled in DSE. Lastly, because of the commit ordering rule, \sysname performs best when communicating services persist at roughly the same pace (e.g., every 10ms), so it is less likely to delay messages or force persistence. In cases where services need to be out of sync, additional mechanisms are required -- for example, one might allow recipient of a message with persistence number difference bounded by a fixed number or ratio. We leave more detailed exploration of these topics for future work.

%% file: sections/eval.tex
\section{Evaluation}
\label{sec:eval}
Our evaluation studies the following research questions:
\begin{squishitemize}
  \item Does DSE help applications reduce end-to-end latency?
  \item What overhead does \sysname impose on applications?
  \item Does \sysname scale to large numbers of participants or high concurrency on a single service?
\end{squishitemize}

\textbf{Experimental Setup.}
We run all of our experiments on the Azure public cloud. For end-to-end experiments, we build the applications using the popular gRPC + ASP.NET stack and deploy them onto Kubernetes cluster. We use a managed Azure Kubernetes Service (AKS) cluster~\cite{aks}; all workloads are scheduled onto a pool of 10 Standard\_D8s\_v3 machines~\cite{dv3series} with attached premium locally redundant storage SSDs~\cite{plrs}. Unless otherwise specified, all \sysname services run with a group commit frequency of 10ms. For microbenchmarks, we use a pair of D32s\_v3 machines~\cite{dv3series} (as client and server), each with 32 vCPUs and 128 GB of RAM. 

\subsection{End-to-End Benchmarks}

\begin{figure}[t!]
    \centering
    \setlength{\fboxrule}{0pt}
    \fbox{
        \includegraphics[width=\columnwidth]{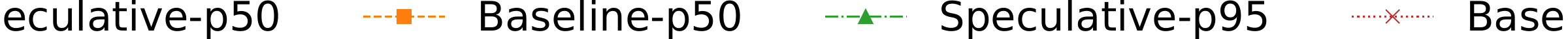}
    }
    \\
    \hfill
    \subfloat[Varying \#Services]{
      \includegraphics[width=0.45\columnwidth]{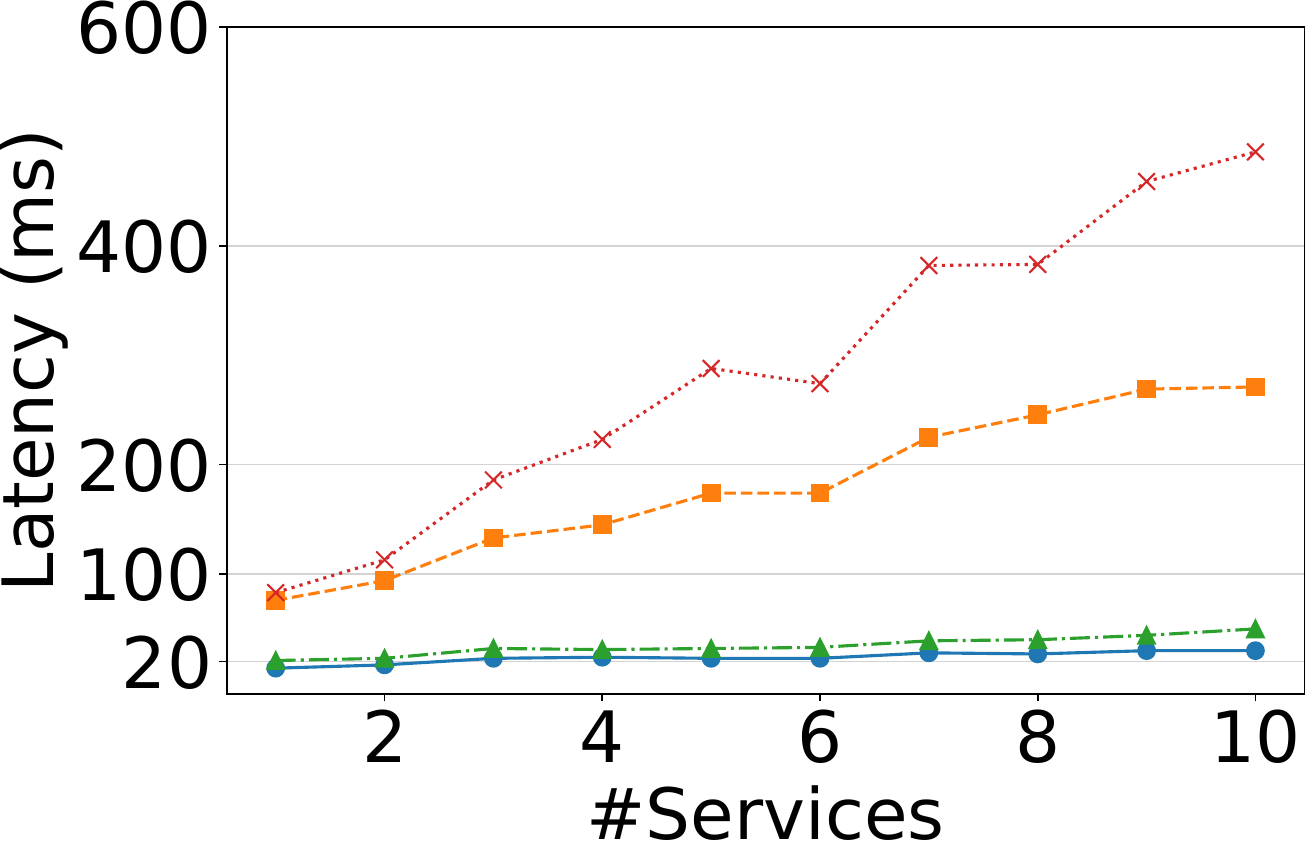}
    }
    \hfill
    \subfloat[Varying Load]{
      \includegraphics[width=0.45\columnwidth]{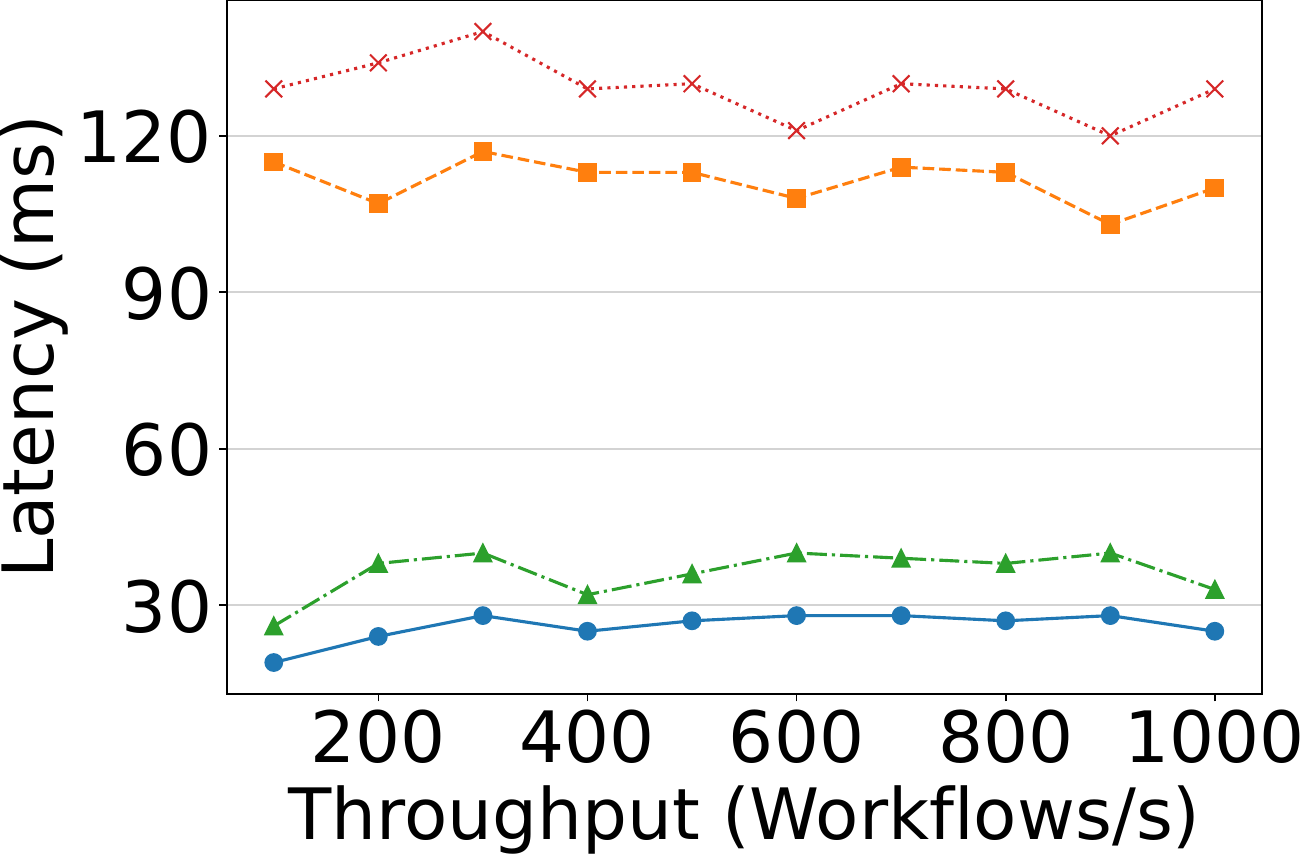}
    }\vspace{-1em}
    \caption{
        \textbf{TravelReservations}
    }
    \label{fig:travel-reservations}
\end{figure}

\textbf{TravelReservations.} We first assemble a travel reservation system based on DeathStarBench~\cite{deathstarbench}. Similar to previous work~\cite{zhang2020beldi, jia21boki, zhuang23exoflow}, we focus on the \emph{write} portion of the workload, where the speculative workflow engine reserves a series of items, one from each service (e.g., hotel, flight, car-rental) backed by our speculative KV store using sagas~\cite{sagas}. 
Because prior systems such as Beldi and Boki are built on different stacks (golang + AWS + Lambda), our system is not directly comparable with them. Instead, as a baseline, we simulate systems such as Beldi and Boki by turning off speculation in our system; doing so will cause our system to perform the same number of synchronous persistence as Beldi/Boki would, but normalize performance of other parts of the system so they are comparable with our implementation. All benchmarks run for 120 seconds. \cref{fig:travel-reservations} shows the result of our experiment. On the left we report the average and 95th-percentile latency when varying the number of services involved, while issuing at a low rate of 10 workflows per second for accurate latency measurements. When the number of services increase, the latency of the baseline linearly increases due to synchronous persistence. Our DSE implementation, on the other hand, experiences only minor latency increase stemming only from increased RPC work instead of synchronous persistence. 
We further demonstrate that DSE has no detrimental effects on scalability by fixing the number of services to 3, and varying the arrival rate of workflows. As shown on the right side of \cref{fig:travel-reservations}, both versions of our implementation scale up to thousands of workflows without performance degradation, similar with numbers reported by Beldi, Boki, and ExoFlow. 
\begin{figure}[t!]
    \centering
    \setlength{\fboxrule}{0pt}
    \fbox{
        \includegraphics[width=\columnwidth]{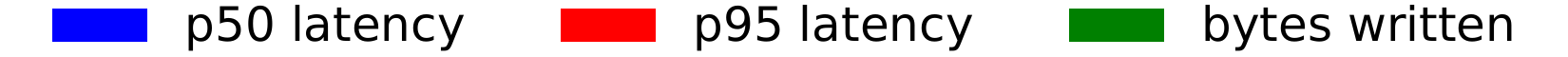}
    }
    \\
    \hfill
    \subfloat[c=10ms]{
      \includegraphics[width=0.4\columnwidth]{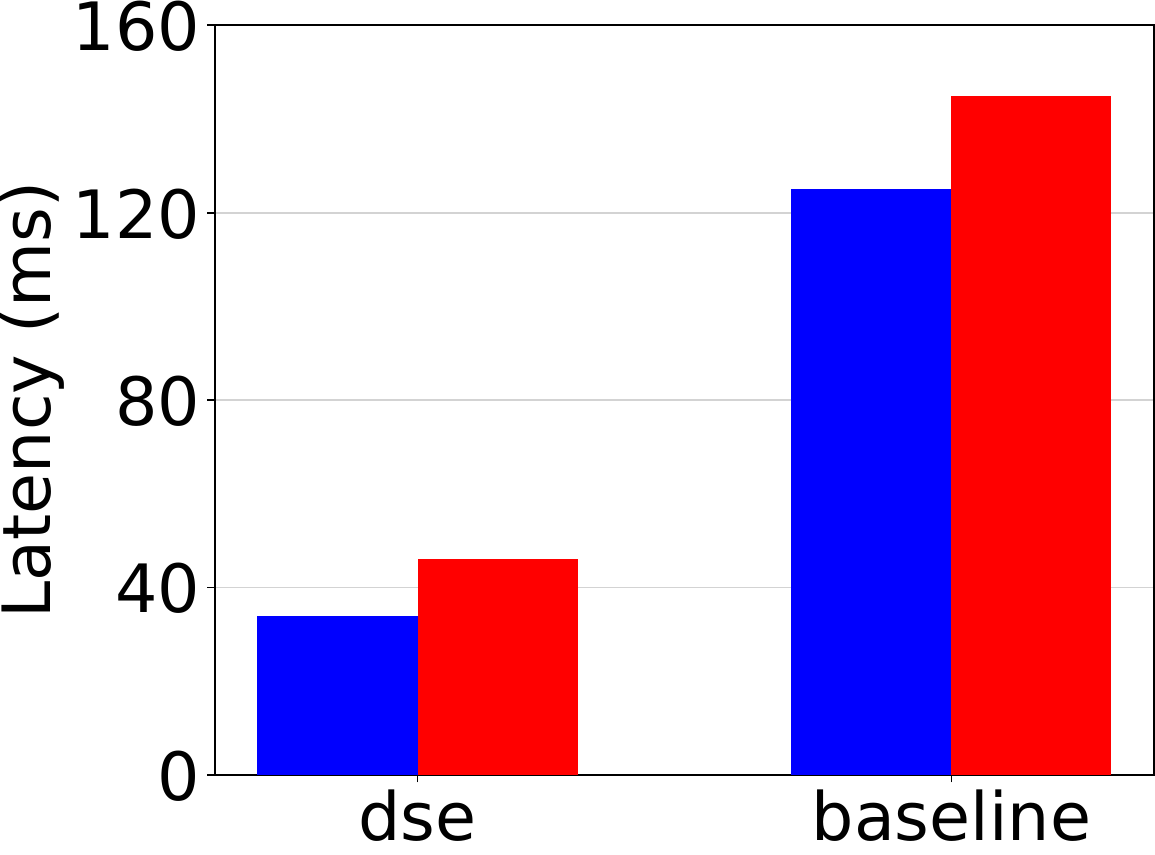}
    }
    \hfill
    \subfloat[c=500ms]{
      \includegraphics[width=0.4\columnwidth]{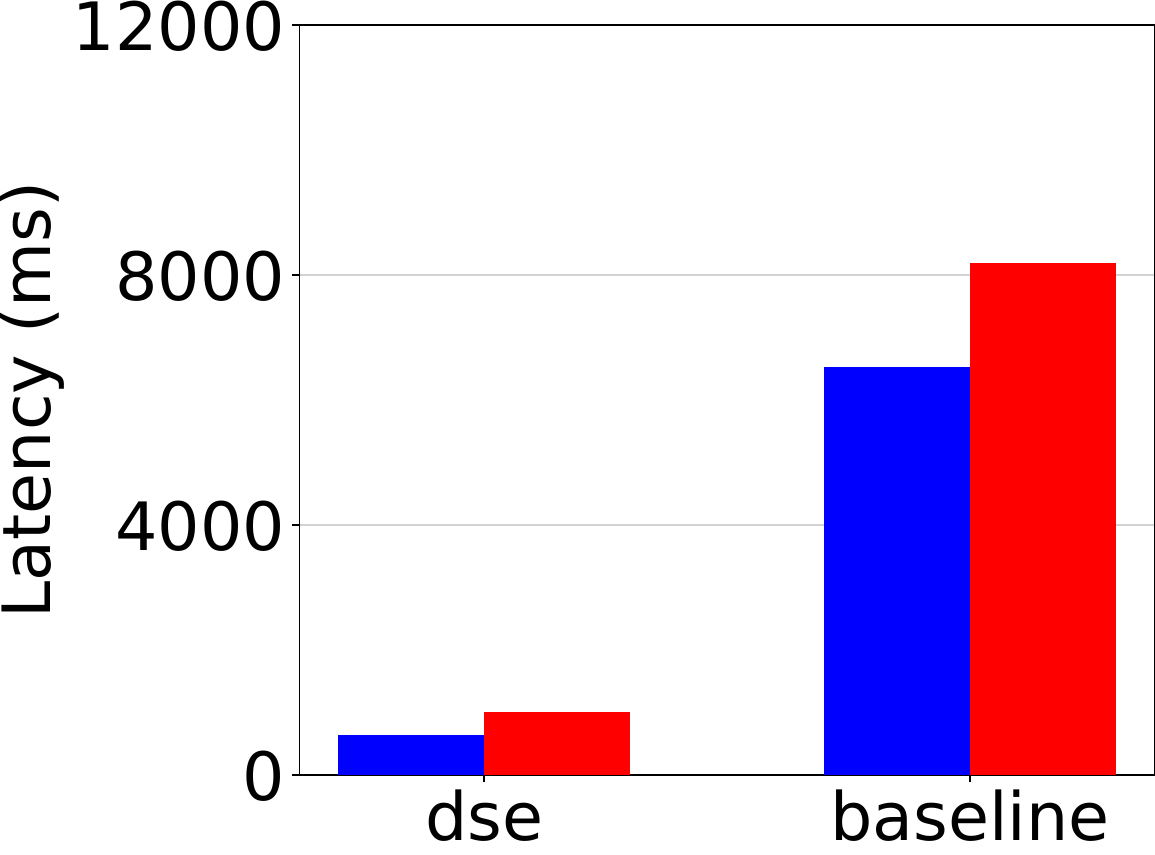}
    }
    \hfill
    \subfloat[bytes written]{
      \includegraphics[width=0.65\columnwidth]{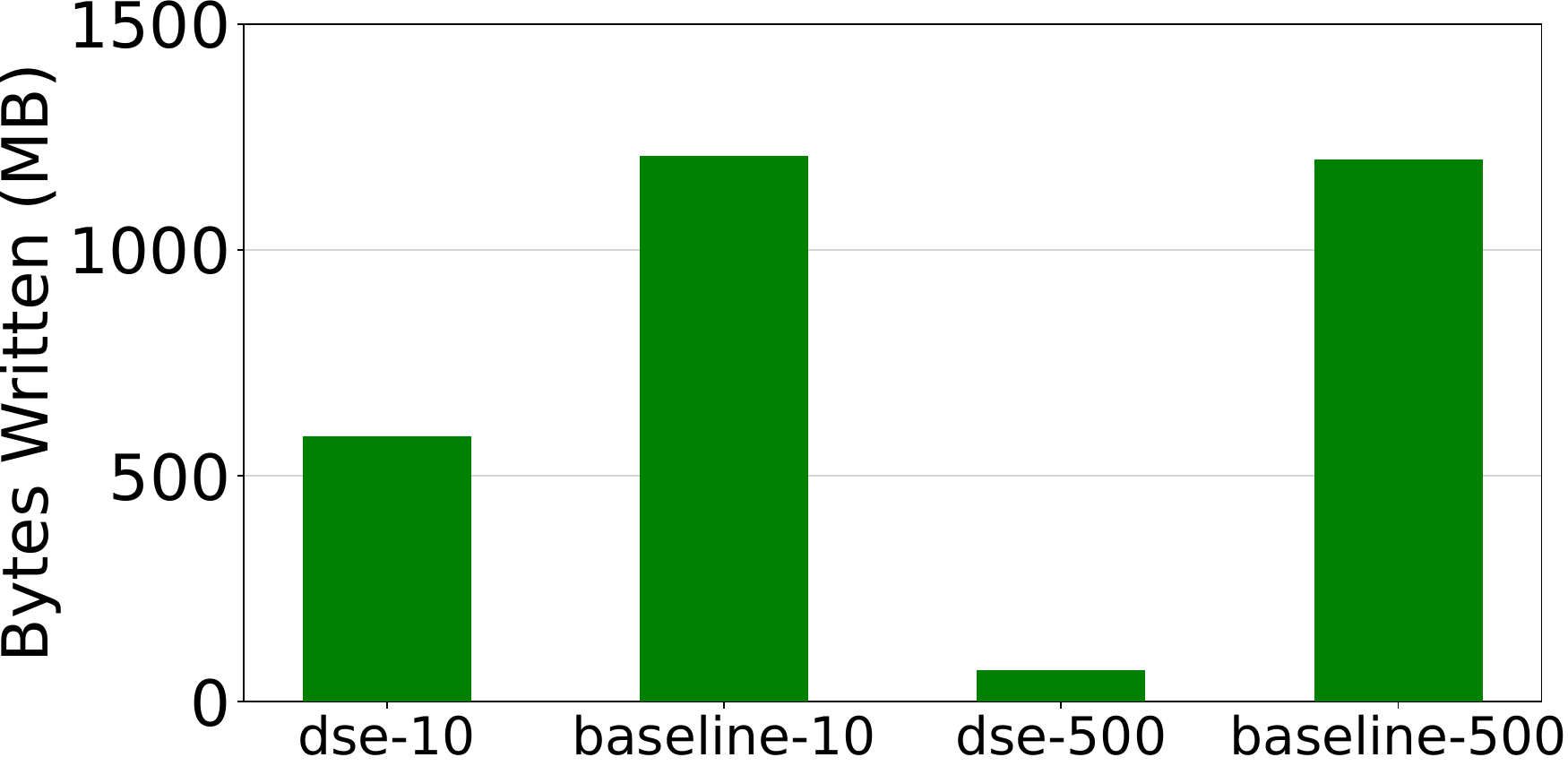}
    }
    \caption{
        \textbf{EventProcessing}}
    \label{fig:event-processing}
\end{figure}

\textbf{Event Processing.} We use our speculative event broker to implement the the search trend alert workload as described in the DARQ paper\cite{li23darq} (available in open-source), with three processing stages communicating with each other through events in a streaming fashion. 
We compare our implementation with the original, non-speculative version of DARQ. \cref{fig:event-processing} shows the result of our experiment. We issue a pre-generated trace of events at a steady rate of 50k events/s for 120 seconds, varying the group commit frequency ($c$).  The first two plots demonstrate latency savings, where DSE drastically reduces the end-to-end latency of the workload. Importantly, in the case of stream processing, DSE is more than a latency saving; the bottom graph shows the number of bytes written to storage during execution, and as shown, DSE offers significant reductions in storage bandwidth. This is because, as an optimization, if an intermediate processing result is generated, consumed, and pruned from the system during speculative execution, it never needs to reach storage. Hence, as shown in \cref{fig:event-processing}, this storage saving is more profound with a large group commit period, which increases the probability that intermediate results never reach storage.

\begin{figure}[t!]
    \centering
        \hfill
    \subfloat[speculative]{
      \includegraphics[width=0.45\columnwidth]{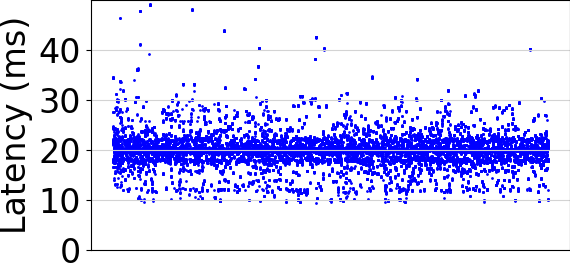}
    }
    \hfill
    \subfloat[non-speculative]{
      \includegraphics[width=0.45\columnwidth]{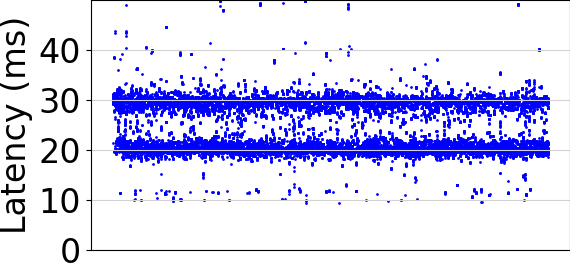}
    }

    \caption{
        \textbf{TwoPhaseCommit}
    }
    \label{fig:twopc}
\end{figure}

\textbf{Two Phase Commit.} Finally, we build a textbook two-phase commit protocol~\cite{mohan86transaction} using \sysname. The benchmark cluster consists of one commit coordinator and four commit participants, each writing to their own (speculative, when speculation is enabled) log. Each transactional client first (without waiting for persistence) writes a log record to each participant signaling the start of a transaction, and then issues a request to the coordinator to begin commit. Participants always vote yes unless the commit start record has been lost due to a failure. Note that our implementation only contains the commit logic and does not include any actual data operation. We issue 30000 transactions from 8 concurrent closed-loop clients to the coordinator, and plot the commit latency of each transaction in \cref{fig:twopc}. As seen, in the non-speculative version, commit latency clusters around multiples of 10, which is the group commit frequency. There are a few ``lucky'' transactions that  complete the first round of messages at the end of the last group commit and therefore finish close to 10ms, but most transactions wait for longer. With speculative execution, the distribution becomes less bimodal, as prepare messages no longer wait for commit, and much more frequently, commits finish under 20ms (i.e. do not wait for two group commits).
This leads to average latency savings of about 5ms. 

\subsection{Recovery}
\begin{figure}[t!]
    \centering
    \hfill
    \subfloat[speculative]{
      \includegraphics[width=0.45\columnwidth]{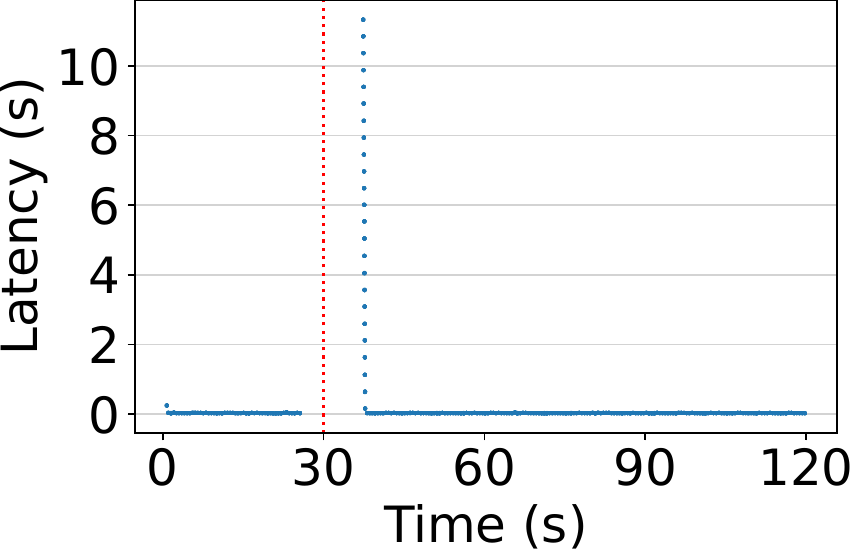}
    }
    \hfill
    \subfloat[non-speculative]{
      \includegraphics[width=0.45\columnwidth]{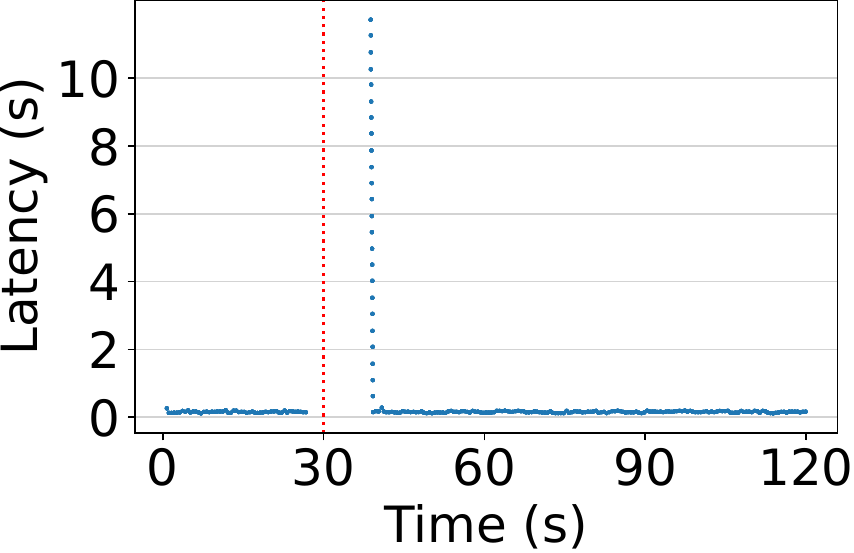}
    }

    \caption{
        \textbf{EventProcessing-recovery}
    }
    \label{fig:event-processing-recovery}
\end{figure}
\textbf{Event Processing.} We first show realistic recovery by running the EventProcessing workload from the previous section and killing one of the event handling nodes in kubernetes at the 30 second mark. Kubernetes then automatically restarts the failed node on a new container, which triggers \sysname to recover when the new node starts. Processing is temporarily halted to the failed node as gRPC requests to it timeout and are retried. Recall that compared to non-speculative applications, speculative applications additionally perform rollback recovery after the failed node restarts. As seen in \cref{fig:event-processing-recovery}, both versions of the application takes around 10 seconds to recover, which mostly stems from container restart. The overhead of speculative recovery is negligible compared to restart. 

\begin{figure}[t!]
    \centering
    \hfill
    \subfloat[speculative]{
      \includegraphics[width=0.45\columnwidth]{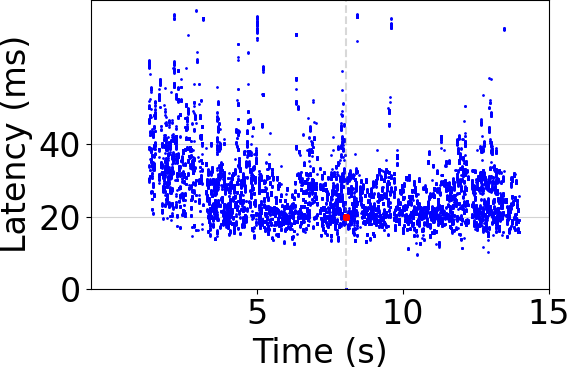}
    }
    \hfill
    \subfloat[non-speculative]{
      \includegraphics[width=0.45\columnwidth]{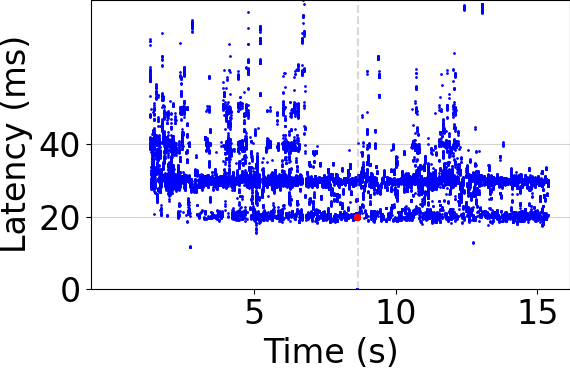}
    }

    \caption{
        \textbf{TwoPhaseCommit-recovery }
    }
    \label{fig:two-pc-recovery}
\end{figure}
\textbf{Two Phase Commit.} To further study the effect of speculation on recovery performance, we construct a recovery scenario based on the two-phase commit workload. We force one of the transaction workers to atomically rollback to a previously committed state, losing all uncommitted state, simulating an instantaneous fail-over to a stand-by. This represents the theoretical best case recovery performance, which leaves only recovery overhead induced by speculative execution. To increase the complexity of recovery, we increase the number of concucrrent clients to 64 in this experiment. We plot the results in \cref{fig:two-pc-recovery}, where red dots represent aborted transactions. Firstly, we observe that in both cases, recovery is very fast and does not create a noticeable gap in operations, except through aborted operations. In comparison, however, speculative recovery results in more aborts. This is because aborts can only come from committers losing transaction start records in the non-speculative case, but \sysname will aggressively rollback any potentially inconsistent state, aborting more transactions than necessary.

To summarize, even though DSE makes recovery more expensive and complex, and could aggressively rollback more work than necessary, it does not negatively impact applications because recovery itself is rare and already expensive. 

\subsection{Microbenchmarks}
Finally, we present a series of microbenchmarks to study the overhead of various \sysname mechanisms. 
\begin{figure}[t!]
    \centering
    \setlength{\fboxrule}{0pt}
    \fbox{
        \includegraphics[width=\columnwidth]{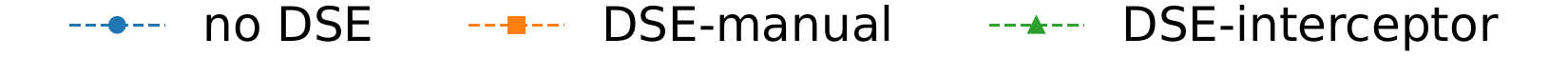}
    }
    \includegraphics[width=0.6\columnwidth]{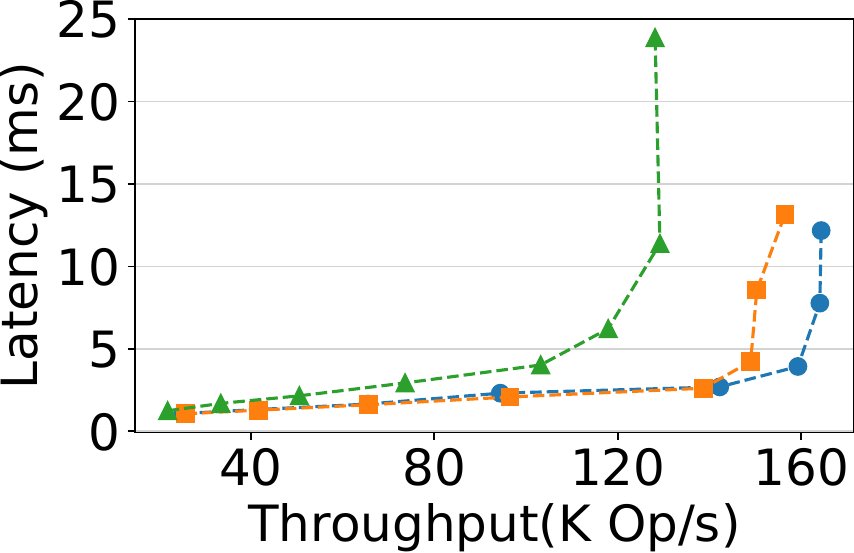}
    \caption{
        \textbf{Overhead of \sysname instrumentation}
    }
    \label{fig:dse-faster}
\end{figure}

\textbf{Overhead of \sysname Instrumentation.} We first study the latency and throughput overhead of \sysname's message instrumentation protocol on a service. We study three versions of the FASTER-based key-value store from the hotel reservation benchmark, two with \sysname (DSE) and another as a thin RPC wrapper around the vanilla FASTER code (no DSE). To study the effects of the interceptor mechanisms vs the instrumentation protocol itself, we also build one version that process \sysname headers in user code (DSE-manual) and another one that does so transparently with gRPC interceptors (DSE-interceptor). We vary request issue rate for each configuratio nto investigate the resulting latency-throughput trade-off. As seen in \cref{fig:dse-faster}, \sysname with interceptors results in slightly higher latency when the system is not saturated and around 25\% lower maximum throughput. This is mostly due to the additional work entailed by the gRPC interceptor mechanism (e.g., HTTP header manipulation). In contrast, the \sysname protocol itself causes negligible increase in latency and less than 5\% reduction in throughput. This shows that the \sysname instrumentation adds only a small overhead to applications.

\begin{figure}[t!]
    \centering
    \hfill
    \subfloat[local-action]{
      \includegraphics[width=0.3\columnwidth]{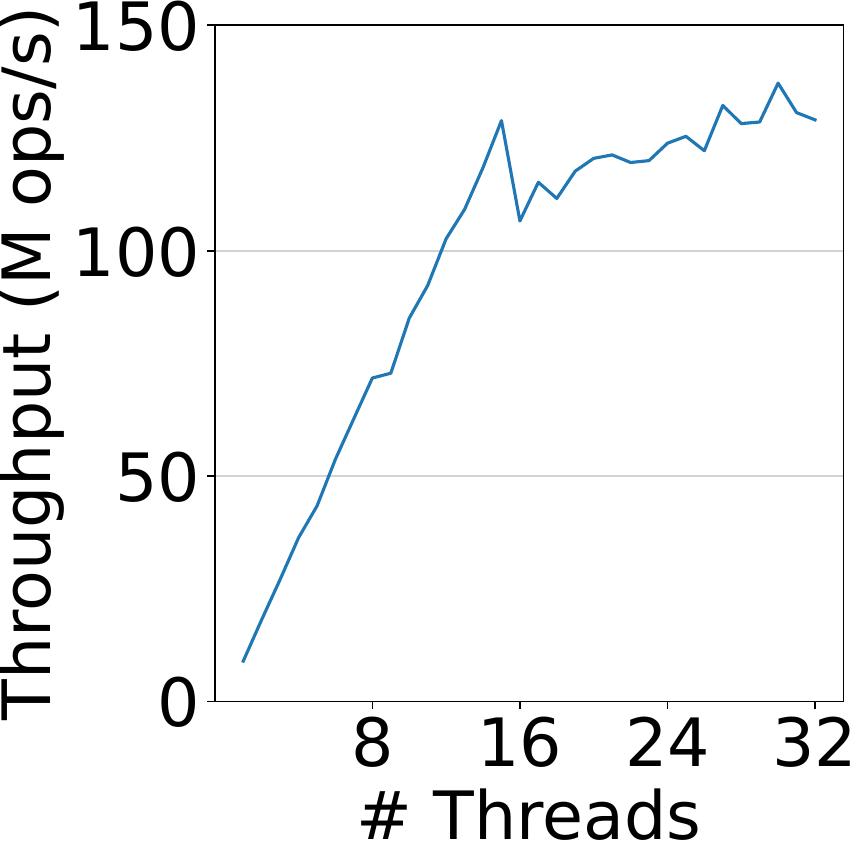}
    }
    \hfill
    \subfloat[send-receive]{
      \includegraphics[width=0.3\columnwidth]{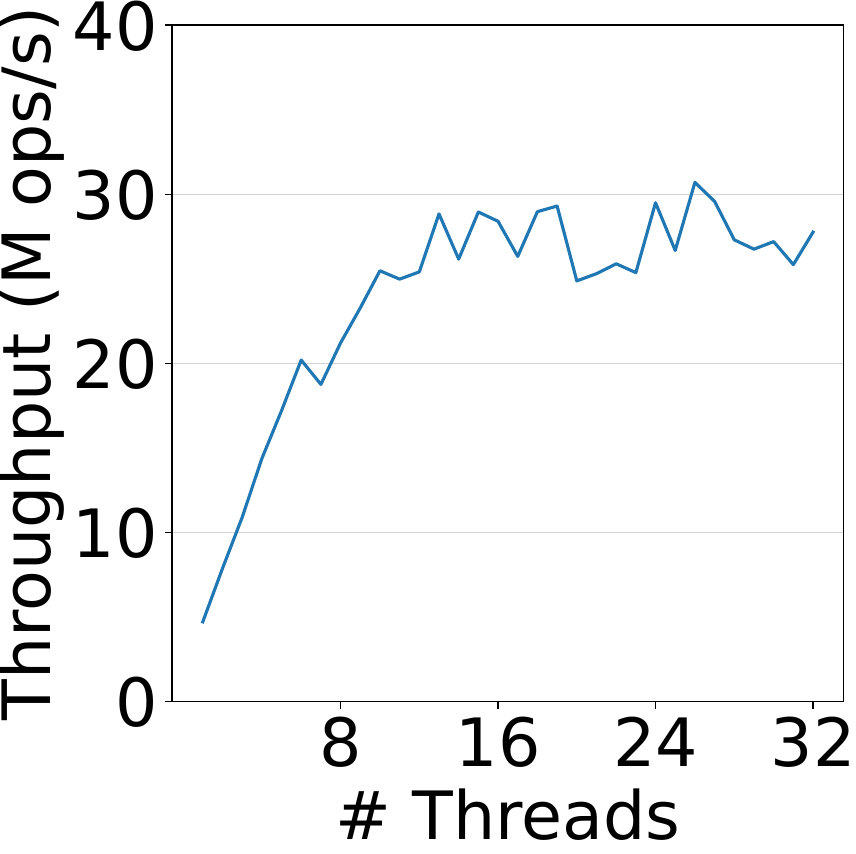}
    }
    \hfill
    \subfloat[detach-merge]{
      \includegraphics[width=0.3\columnwidth]{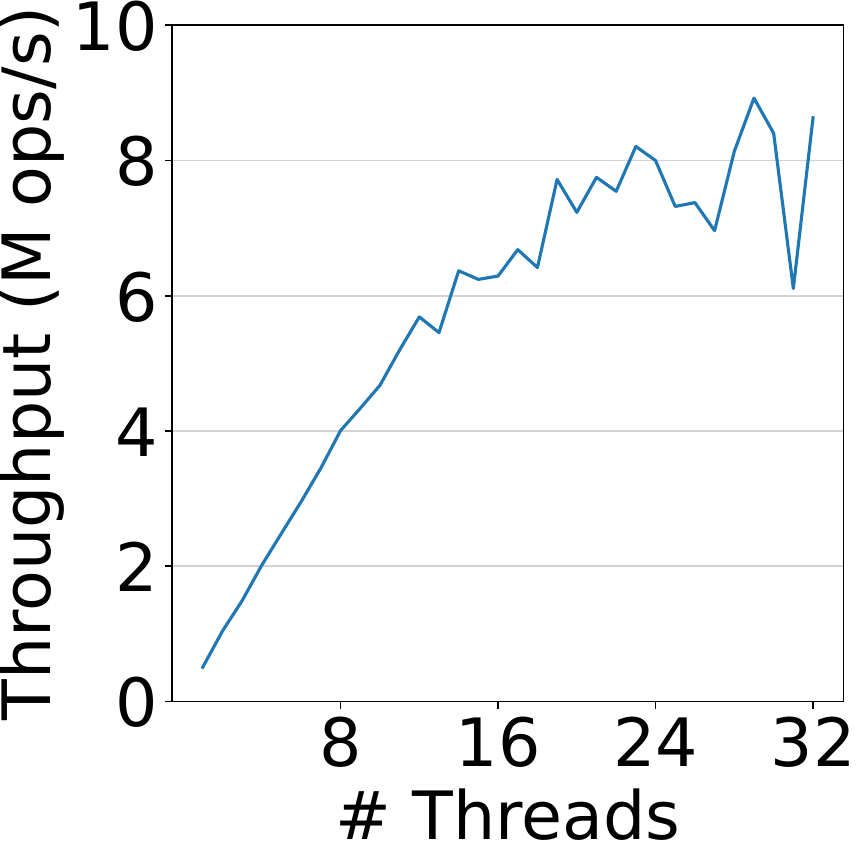}
    }
    \caption{
        \textbf{Thread scalability of \sysname Primitives}
    }
    \label{fig:thread-scalability}
\end{figure}
\textbf{Thread Scalability of \sysname Primitives.} Recall that \sysname users must protect their operations with \sysname threading primitives for correct speculative execution. It is therefore important the \sysname primitives be scalable so as to not introduce new bottlenecks in an application. To measure this, we design a simple microbenchmark where threads acquire protection under a tight-loop concurrently, reporting throughput as the number of concurrent threads increase. A background thread periodically performs (empty) checkpoints to advance versions. We measure three sets of primitives: local-action, where \sysname users mark blocks of code blocks as atomic with regards to checkpointing and recovery, send-receive, where \sysname users start an action using a (randomly pre-populated) header and writes the resulting dependencies to another header at the end of action, and finally detach-merge, where \sysname users create a new \session from the current dependencies and then merging it with the state object. As shown in \cref{fig:thread-scalability}, our implementation of the \action primitive (based on \cite{li22epvs}) is scalable up to 16 threads with no adverse degradation of performance. As expected, send-receive is more expensive then local-action, and detach-merge than send-receive, due to the amount of computation involved. However, all three primitives can sustain millions of operations per second on a single server, making it unlikely to be a bottleneck in typical RPC-oriented workloads that only see 100s of thousands of operations per second per node. 

%% file: sections/rel.tex
\section{Related Work}
\label{sec:rel}
  
\noindent\textbf{Durable Execution and Workflows.} Durable Execution, as a term, is first coined by Temporal~\cite{durable-execution}, and widely accepted in industry~\cite{riccomini23, restate, orkes, stealthrocket, littlehorse, flawless, convex, rama}, although it is similar to the earlier concept of virtual resiliency~\cite{ambrosia, goldstein2018ambrosia}.
Workflow systems~\cite{airflow, cloudcomposer, kubeflow}, orchestrate execution of a directed acyclic graph (DAG) of tasks, and is widely used in industry to compose complex applications from individual components. Workflow systems are some of the earliest adopters of durable execution, providing exactly-once guarantees over the DAG. 
Systems such as Azure Durable Functions~\cite{df}, and Temporal~\cite{temporal} achieves this by synchronously persisting workflow state before starting new steps; external effects of execution are wrapped in special abstractions (e.g., activities in Temporal and DF) and required to be idempotent for resilience. ExoFlow~\cite{zhuang23exoflow} is a recent workflow system that allows developers to take advantage of determinism and rollbacks within the workflow for better performance. Developers mark parts of their workflow as deterministic or rollback-capable, and ExoFlow will bypass synchronous persistence where possible, using replay and rollback to recover when necessary. In contrast, \sysname works for any message-passing fail-restart application, and does not rely on determinism.

\noindent\textbf{Cross-Service Transactions.} Other systems, such as Olive~\cite{setty16olive}, co-locate service state and orchestration state onto one unified storage layer, providing exactly-once guarantees without requiring idempotence, but also forces developers to migrate to their storage engine. While Olive used existing database services (e.g., Azure Tables), Boki~\cite{jia21boki} showed that a shared log abstraction can provide more performance and usability benefits, and Halfmoon~\cite{qi23halfmoon} further optimizes Boki to reduce the amount of information logged. One advantage of this architecture is that the unified storage layer can be engineered to support distributed transactions~\cite{zhang2020beldi, sreekanti20aft}. Orleans, of particular note, implemented early lock release~\cite{gawlick85, dewitt84, eldeeb2016transactions, berstein19orleans, eldeeb24} to reduce latency, which is similar to DSE. In contrast, \sysname is designed to work for heterogeneous storage engines, and do not implement transactional semantics (transactions can optionally be layered on top). While some protocols allow distributed transactions across heteregenous storage engines~\cite{xa, kraft23epoxy, zhang22skeena}, they are not widely used currently to our knowledge. Finally, DBOS~\cite{dbos, kraft2023apiary} proposes to build all services and runtime layers on top of distributed transactional DBMS, but would require an even more radical rewrite of existing technology stacks than \sysname. 

\noindent\textbf{Other Related Work.} Fault-tolerance in microservice-oriented cloud applications can be viewed as a modern instantiation of asynchronous recovery in message-passing systems, which was heavily studied in prior work~\cite{elnozahy02survey}. 
The programming model of \sysname is similar in concept to actor systems~\cite{orleans}, where distributed stateful objects communicate through asynchronous method calls, but usually without resilience guarantees.
Dataflow systems are conceptually similar to workflow systems and compose distributed components, but usually have the benefit of known application semantics (i.e, composed components are operators such as filter and map, rather than blackbox code). Consequently, dataflow systems predominantly use lineage-based techniques to reconstruct application state on failure~\cite{zaharia12resilient, murray13naiad, wang19lineage}. \sysname is, at its core, a checkpoint-based recovery scheme, which assumes no such domain-specific knowledge. Finally, Speculator~\cite{nightingale05speculator} and xsyncfs~\cite{nightingale08rethink} similarly uses speculative execution to speed up file system operations transparently. However, they are different from DSE in two key aspects. Firstly, speculator and xsyncfs are \emph{client-centric}: each client process is responsible for checkpointing their own state, blocking operations that externalize speculative state, and rolling back on failed speculation. File servers never store speculative state. In contrast, in DSE, distributed participants collaborate on speculative state and coordinate to repair state after failure. Secondly, DSE application only speculate on whether operations are lost (i.e., wait for the operation to complete, but not persistence), whereas speculator attempts to speculate on the result (i.e., do not wait for the operation to complete, but guess the result). This is because failed speculation is cheap for speculator, as rollbacks are localized to client processes. DSE rollbacks, in contrast, involve distributed coordination among many affected components, and should only be triggered when necessary (i.e., when an actual failure occurs).